\documentclass[%
 reprint,
 amsmath,amssymb,
 aps,
]{revtex4-2}

\usepackage{graphicx}% Include figure files
\usepackage{dcolumn}% Align table columns on decimal point
\usepackage{bm}% bold math
\begin{document}

\preprint{APS/123-QED}

\title{Minimum energy paths for defect annihilation in finite dodecagonal quasicrystal clusters}

\author{Tiejun Zhou}
\affiliation{Hunan Key Laboratory for Computation and Simulation in Science and Engineering,
Key Laboratory of Intelligent Computing and Information Processing of Ministry of Education,
School of Mathematics and Computational Science,
Xiangtan University, Xiangtan, Hunan 411105, China}

\author{Rong Liu}
\affiliation{Hunan Key Laboratory for Computation and Simulation in Science and Engineering,
Key Laboratory of Intelligent Computing and Information Processing of Ministry of Education,
School of Mathematics and Computational Science,
Xiangtan University, Xiangtan, Hunan 411105, China}

\author{Gang Cui}
\affiliation{Wuhan Institute for Math \& AI,
Wuhan University, Wuhan, Hubei 430072, China}

\author{Xianbo Huang}
\affiliation{National-certified Enterprise Technology Center,
Kingfa Science and Technology Co., Ltd.,
Guangzhou, Guangdong 510663, China}

\author{Li Peng}
\affiliation{National-certified Enterprise Technology Center,
Kingfa Science and Technology Co., Ltd.,
Guangzhou, Guangdong 510663, China}

\author{Kai Jiang}
\email{kaijiang@whu.edu.cn}
\affiliation{Wuhan Institute for Math \& AI,
Wuhan University, Wuhan, Hubei 430072, China}

\collaboration{CLEO Collaboration}%\noaffiliation

\date{\today}% It is always \today, today,
             %  but any date may be explicitly specified

\begin{abstract}
Defect repair in quasicrystals involves local structural
rearrangements on a complex potential energy landscape.
Here, we combine the string method with the spring pair method
to compute a minimum energy path connecting a locally defective
finite dodecagonal quasicrystal cluster to a locally repaired
reference configuration in the Lennard-Jones--Gauss particle
model. The computed path contains seven local minima connected
through six transition states. A detailed analysis of one
defective region resolves three successive stages: a phason
flip, aggregation of neighboring shield-like defects, and
decomposition of the aggregated motif. The phason flip restores
the local outer-ring structure and relocates a shield-like
defect into the interior. Subsequent collective rearrangements
remove the defective motifs and yield a net decrease in the
local interaction-energy measure together with an increase
in local bond-orientational order. These results provide a
microscopic description of a specific defect-repair pathway
on the potential energy surface of a finite dodecagonal
quasicrystal cluster.
\end{abstract}

%\keywords{Suggested keywords}%Use showkeys class option if keyword
                              %display desired
\maketitle

%\tableofcontents
\section{Introduction}

Quasicrystals (QCs) exhibit long-range orientational and positional order without translational periodicity. They have attracted significant interest in materials science and condensed matter physics since their discovery\cite{shechtman1984metallic,levine1984quasicrystals,lieu2022formation,wang2025defect}. In mathematical descriptions, ideal QCs can be constructed through higher-dimensional projections, matching rules, or inflation tilings. In physical systems, quasicrystalline order can coexist with local structural defects. For square–triangle dodecagonal QCs, such defects include shield-like motifs and associated local tiling irregularities, identified from their particle arrangements and connectivity relative to the regular square-–triangle tiling\cite{ulugol2025defects,alptug2026vacancy}. Therefore, we use the term \emph{defect quasicrystal} to refer to finite quasicrystal clusters containing local geometric structural defects, whereas \emph{defect-repaired quasicrystal} refers to finite quasicrystal clusters in which such geometric structural defects have been repaired.

Defect quasicrystals have been widely reported in metal alloys and soft matter. For example, icosahedral QCs in the Khatyrka meteorite contain minor defects\cite{bindi2016collisions}. Experimentally synthesized dodecagonal QCs, such as Ni--Cr alloys and silica, frequently exhibit shield-like defects\cite{ishimasa1988electron,ishimasa2011dodecagonal,xiao2012dodecagonal}. Various types of local defects are also found in colloids and liquid crystals\cite{ulugol2025defects,dontabhaktuni2018quasicrystalline,kim2022quasicrystalline}. In square--triangle dodecagonal systems, shield-like defects, egg-like defects, bowtie-like motifs, and vacancy-related structures are representative local configurations associated with phason flips and tile rearrangements\cite{oxborrow1993random,ulugol2025defects,alptug2026vacancy,zhao2025atomistic}. The relaxation or annihilation of such local defects can improve local structural order and influence defect mobility, local stability, and the mechanical response of quasicrystalline particle systems\cite{engel2007stability,engel2011entropic,huttunen2004microstructure,mbah2023early,beardsley2008potential,zhou2004study}. Therefore, elucidating the microscopic pathways of defect annihilation is important for understanding local relaxation processes and defect-controlled structural evolution in quasicrystalline systems.

Over the past decades, defect annihilation and local rearrangements in quasicrystals have been widely investigated. Experimentally, phason-flip-mediated annihilation of bowtie- and star-shaped defects has been observed in decagonal $\mathrm{Al_{74}Cr_{15}Fe_{11}}$\cite{ma2020novel}, and growth studies of decagonal Al--Ni--Co QCs have shown local recovery of quasiperiodic order after transient deviations\cite{nagao2015experimental}. In simulations, Brownian dynamics and molecular dynamics have clarified the roles of defect mobility, phason flips, zipper-like rearrangements, and shield-like defect dynamics in quasicrystalline particle systems\cite{achim2014growth,schmiedeberg2017dislocation,gemeinhardt2018growth,zhao2025atomistic,engel2010dynamics}. In particular, the geometric mechanisms underlying tile rearrangement in the square-triangle-dodecagon system have been studied in depth\cite{oxborrow1993random,ulugol2025defects}. Nevertheless, resolving the complete minimum energy path (MEP) of defect repair in a particle-level quasicrystalline model remains challenging because the potential energy surface (PES) is high-dimensional and rugged. Although recent nudged elastic band calculations have identified local defect migration pathways in the Lennard-Jones--Gauss (LJG) system with periodic boundary conditions\cite{zhao2025atomistic}, a saddle-resolved description of how prescribed shield-like defects relax into a locally repaired finite dodecagonal cluster remains incomplete. Such a pathway is needed to quantify intermediate minima, index-1 saddles, energy barriers, and particle-level rearrangements during local defect repair.

In this work, we study the structural transformation from defect dodecagonal quasicrystals (de-DDQC) to defect-repaired dodecagonal quasicrystals (dr-DDQC). Using the LJG model, we consider a finite-size quasicrystal cluster under free boundary conditions, where particles interact solely through LJG interactions without any confining box or external constraints. We calculate the MEP between de-DDQC and dr-DDQC by combining the string method and the spring pair method\cite{weinan2002string,ren2007simplified,cui2024spring,engel2007self,moritz2024confirmation}. The computed MEP resolves a zero-temperature energetic pathway for local defect repair. We find that the defect-repair process follows three stages: phason flip, aggregation of shield-like defects, and decomposition of the aggregated shield-like defect. A phason flip facilitates the formation of the outer ring of the cluster via the motion of a single particle whose energy remains nearly unchanged. Furthermore, the aggregation and decomposition of shield-like defects are driven by local potential-energy gradients and involve the coordinated motion of multiple particles, ultimately leading to the local annihilation of the prescribed defects. These transformations together provide a saddle-resolved microscopic picture of defect repair in the selected finite LJG dodecagonal quasicrystal cluster.

\section{Model and Methods}
\subsection{Lennard-Jones--Gauss Potential}

We consider a two-dimensional finite cluster of $N=1024$
identical particles with free boundaries. The particles
interact through an isotropic LJG
pair potential. We study transition pathways on the
potential energy surface, without introducing thermal
noise or finite-temperature dynamics. The LJG energy is
\begin{equation}
	V(\widetilde{\bm{r}}^N) =
    \sum_{i=1}^{N-1} \sum_{j=i+1}^{N}
    v(|\bm{r}_i-\bm{r}_j|),
    \label{eq:LJG_N}
\end{equation}
where $\widetilde{\bm{r}}^N=(\bm{r}_1,\bm{r}_2,\dots,\bm{r}_N)$ denotes the configuration of all particles. In this zero-temperature setting, local defect evolution is interpreted as structural relaxation on the potential energy surface (PES), rather than as finite-temperature annealing dynamics.

The pair potential is given by
\begin{equation}\small
		v_{\text{LJG}}(r)
	 	=4\epsilon_0\left[
        \left(\frac{\sigma_0}{r}\right)^{12}
        -\left(\frac{\sigma_0}{r}\right)^6
        \right] 
        +\frac{h}{\sigma_G\sqrt{2\pi}}
	 	\exp\left[-\frac{(r-r_G)^2}{2\sigma_G^2}\right],
	 	\label{eq:LJG}
\end{equation}
Here, $\varepsilon_0$ and $\sigma_0$ set the energy and length
scales of the Lennard-Jones contribution. The Gaussian term
is centered at $r_G$, has width $\sigma_G$, and has amplitude
$h/(\sigma_G\sqrt{2\pi})$. For the negative value of $h$ used
here, this term introduces an additional attractive well\cite{engel2007self,zhao2025atomistic}.

\begin{table}[htbp]
\centering
\caption{Parameter values of the LJG potential.}
\label{tab:parameters}
\begin{tabular*}{\columnwidth}{@{\extracolsep{\fill}}ccccc@{}}
\hline
$\epsilon_0$ & $\sigma_0$ & $h$ & $r_G$ & $\sigma_G^2$ \\
\hline
0.5 & 0.88 & -0.874 & 1.879 & 0.02 \\
\hline
\end{tabular*}
\end{table}

\begin{figure}[htbp]
    \centering
    \includegraphics[width=0.4\textwidth]{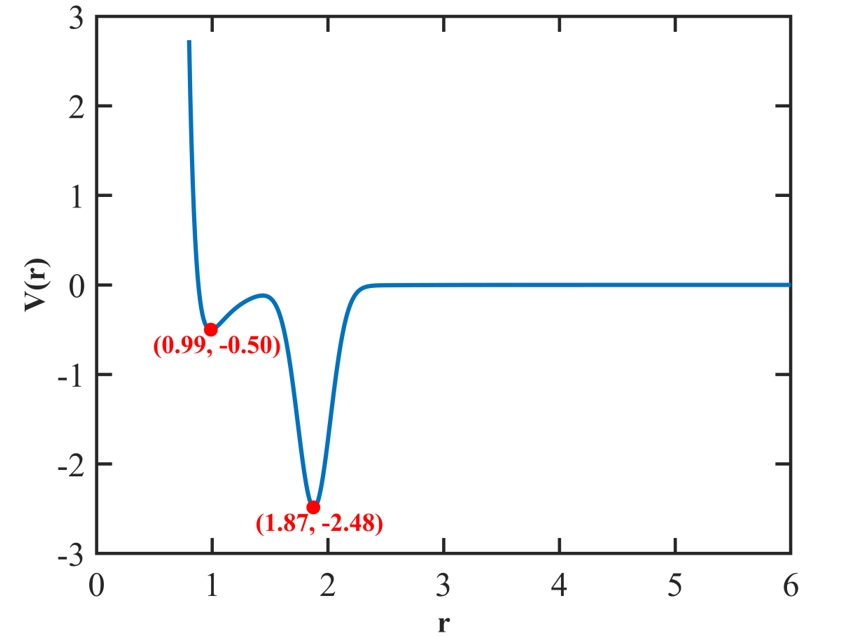}
    \caption{The LJG potential. The two marked minima indicate the competing length scales in the reduced-unit convention used in this work.}
    \label{fig:LJGpotential}
\end{figure}

Following previous simulations of dodecagonal quasicrystals\cite{zhao2025atomistic}, we use the parameter set listed in Table~\ref{tab:parameters}. Unlike standard Lennard-Jones units where the well depth is unity, this parameter set defines a distinct set of LJG-based reduced units: the energy unit is $2\epsilon_0$, and the length unit is related to the first zero-crossing of the potential. All lengths, energies, and derived quantities are reported in these specific LJG-based reduced units.
The resulting double-well profile is shown in Fig.~\ref{fig:LJGpotential}. The two marked minima indicate the competing length scales in the reduced-unit convention used here.

\begin{figure*}[htbp]
    \centering
    \includegraphics[width=1.0\textwidth]{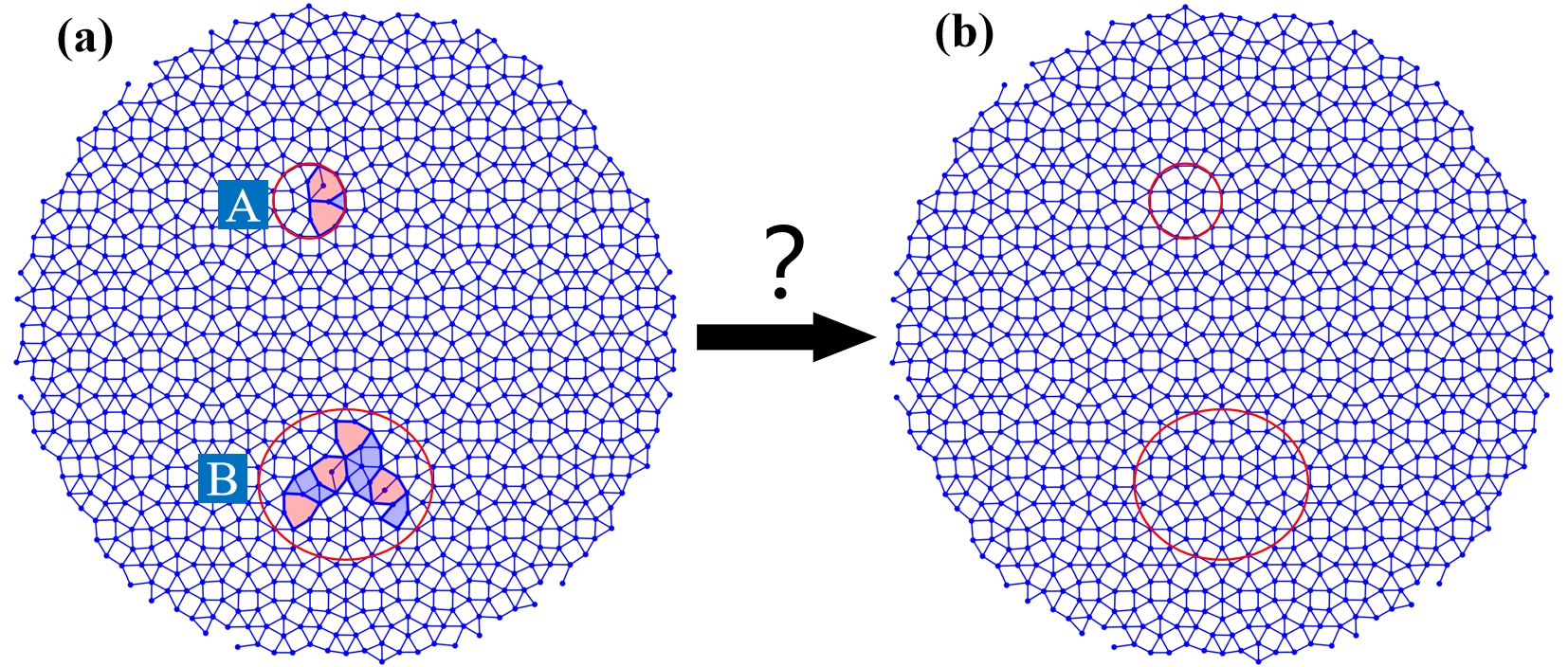}
    \caption{Endpoint configurations for the finite-cluster MEP calculation. (a) de-DDQC, locally defective initial configuration with prescribed shield-like defects. (b) dr-DDQC, locally repaired reference configuration. Red: shield-like defects; blue: secondary local mismatches induced by the defects. A and B indicate upper and lower defective regions.}
    \label{fig:stablestate}
\end{figure*}

The endpoint configurations used in the finite-cluster MEP calculations are shown in Fig.~\ref{fig:stablestate}. Both endpoints correspond to local minima of the total potential energy defined in Eq.~\eqref{eq:LJG_N}. The dr-DDQC configuration is obtained by relaxing a finite quasicrystalline cluster composed of square--triangle tilings. The de-DDQC configuration is constructed by introducing local perturbations into the same cluster, giving rise to shield-like local defects, as shown in Fig.~\ref{fig:stablestate}(a). Although we focus on shield-like defects in the present study, the same MEP-based framework can be readily extended to other local defect motifs, such as egg-like and bowtie-like defects, thereby enabling a systematic investigation of defect-annihilation mechanisms in finite quasicrystal clusters.

\subsection{Saddle Search Method}
To determine the zero-temperature energetic pathway for defect repair, we employ a path-searching strategy to calculate the MEP connecting the de-DDQC and the dr-DDQC. Existing methods for locating saddle points fall into two main categories: path-finding methods and surface-walking methods. Path-finding methods require both the initial and final states to define the endpoints of the pathway, allowing the computation of an MEP on the PES. Along this path, the configuration with the highest energy corresponds to the index-1 saddle point\cite{weinan2002string, ren2007simplified, henkelman2000improved}. Conversely, surface-walking methods start from a single state on the PES and utilize the first or second derivatives of the potential energy to locate index-1 saddle points without requiring a priori knowledge of the final state\cite{weinan2011gentlest, henkelman1999dimer, gould2016dimer, cui2024spring,cui2025efficient,yin2019high}.

Since the complex PES defined by \eqref{eq:LJG_N} is characterized by numerous minima and saddle points, we employ a two-stage strategy combining the string method and the spring pair method (SPM) to compute the MEP between the de-DDQC and dr-DDQC. First, an initial approximate path is efficiently generated using the string method. Subsequently, the SPM is applied to precisely locate the saddle point along this path, thus avoiding the high computational cost associated with full-path optimization.

In the first stage, we approximate the MEP using the string method, which constructs a continuous path discretized into $m$ configuration nodes between the de-DDQC and dr-DDQC. Defining $\bm{F}(\widetilde{\bm{r}}^N) = -\nabla V(\widetilde{\bm{r}}^N) $ as the negative energy gradient (i.e., the force), the evolution equation of the string is given by\cite{weinan2002string, ren2007simplified}
\begin{equation}
	\frac{{\rm d}\widetilde{\bm{r}}^N_k}{{\rm d}t}= \bm{F}(\widetilde{\bm{r}}^N_k) - (\bm{F}(\widetilde{\bm{r}}^N_k) \cdot \bm{\tau}_k) \bm{\tau}_k, \quad k=2,3, \cdots, m-1,
\end{equation}
where $\bm{\tau}_k$ represents the unit tangent vector of the string at node $k$, defined as:
\begin{equation}
	\bm{\tau}_k = \frac{\widetilde{\bm{r}}^N_{k+1} - \widetilde{\bm{r}}^N_{k-1}}{| \widetilde{\bm{r}}^N_{k+1} - \widetilde{\bm{r}}^N_{k-1} |}, \quad k=2, 3, \dots, m-1.
\end{equation}
To prevent excessive node clustering, an equidistant reparameterization is applied after each evolution step, enabling the string to progressively converge toward an approximate MEP\cite{ren2007simplified}.

\begin{figure*}[htbp]
    \centering
    \includegraphics[width=1.0\textwidth]{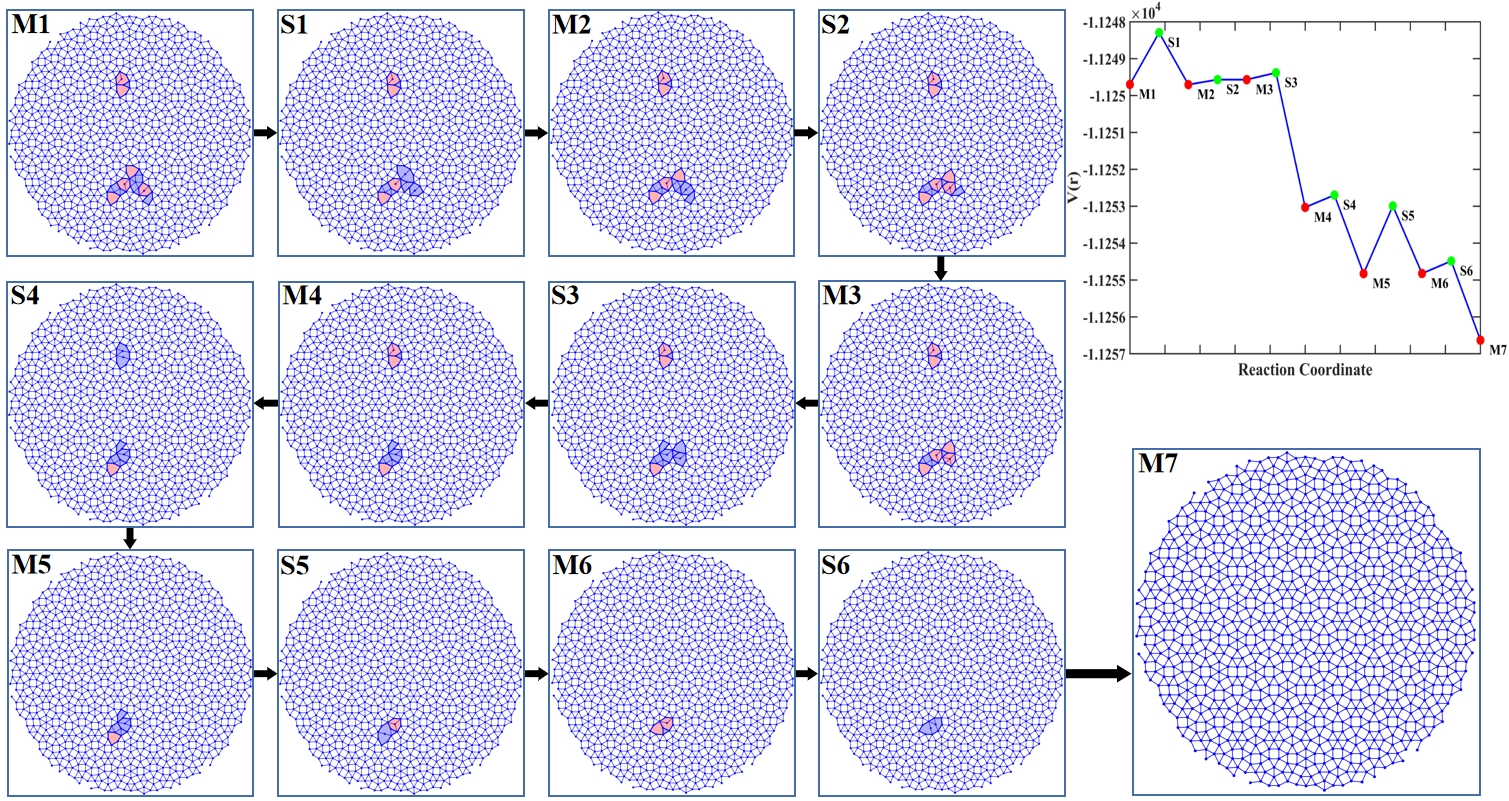} 
    \caption{The transformation pathway from de-DDQC to dr-DDQC. M1 represents the de-DDQC, and M7 represents the dr-DDQC. Mi denotes a minimum and Si denotes an index-1 saddle point. Red-filled marks represent the shield-like defects themselves, and blue-filled indicates secondary defects induced by shield-like defects that do not match the locally repaired quasicrystalline structure. The upper-right panel shows the corresponding energy diagram, where red and green points mark the energies of minima and saddle points, respectively. These states are arranged in pathway order, and their horizontal spacing does not represent physical time.}
    \label{fig:transformation_path}
\end{figure*}

In the second stage, we employ the SPM to precisely locate the saddle point\cite{cui2024spring}, using the highest-energy node from the converged string as an initial guess. The SPM evolves two adjacent configurations, $\widetilde{\bm{r}}^N_1$ and $\widetilde{\bm{r}}^N_2$, which alternately undergo drifting and climbing dynamics to precisely locate the saddle point.

The spring direction vector is defined as $\bm{v}^s = (\widetilde{\bm{r}}^N_2 - \widetilde{\bm{r}}^N_1)/|\widetilde{\bm{r}}^N_2 - \widetilde{\bm{r}}^N_1|$, and the force $\bm{F}(\widetilde{\bm{r}}^N_l)$ on each configuration is decomposed into components parallel and perpendicular to $\bm{v}^s$, 
\begin{equation}
	\begin{cases}
		\bm{F}^{\parallel}(\widetilde{\bm{r}}^N_l)
		= \bigl(\bm{F}(\widetilde{\bm{r}}^N_l)\cdot\bm{v}^s\bigr)\bm{v}^s,\\
		\bm{F}^{\perp}(\widetilde{\bm{r}}^N_l)
		= \bm{F}(\widetilde{\bm{r}}^N_l)-\bm{F}^{\parallel}(\widetilde{\bm{r}}^N_l),
	\end{cases}
	l=1, 2. 
\end{equation}
During the drifting step, the spring pair moves toward the MEP under the action of the perpendicular force component $\bm{F}^{\perp}(\widetilde{\bm{r}}^N_l)$, while the spring force $\bm{F}^{s}(\widetilde{\bm{r}}^N_l)$ adjusts the distance between the two configurations. The evolution in pseudo-time is given by:
\begin{align}
\frac{{\rm d}\widetilde{\bm{r}}^N_l}{{\rm d}t} = \alpha_1 \bm{F}^{\perp}(\widetilde{\bm{r}}^N_l) + \alpha_2 \bm{F}^s(\widetilde{\bm{r}}^N_l), \quad l=1, 2,
\end{align}
where $\alpha_1$ and $\alpha_2$ are algorithmic relaxation parameters.

In the climbing process, where the spring direction $\bm{v}^s$ aligns with the tangent of the MEP, the spring pair is pushed in the direction opposite to the parallel force component $\bm{F}^{\parallel}$ (i.e., up the potential energy gradient) to ascend toward the saddle point:
\begin{align}
\frac{{\rm d}\widetilde{\bm{r}}^N_l}{{\rm d}t} = -\alpha_3 \bm{F}^{\parallel}(\widetilde{\bm{r}}^N_l), \quad l=1, 2,
\end{align}
where $\alpha_3$ is also a relaxation parameter.
Convergence is reached when $\min\left\{|\bm{F}(\widetilde{\bm{r}}^N_1)|,  |\bm{F}(\widetilde{\bm{r}}^N_2)|\right\} < \epsilon$. In this work, we select $\epsilon = 10^{-7}$.

At this stage, the spring orientation approximates the unstable mode. We slightly perturb the system along this unstable direction at the saddle point and follow the steepest-descent paths to both adjacent minima, thereby obtaining a locally refined MEP segment. Because the PES of a finite quasicrystalline cluster is rugged, the resulting path should be interpreted as the MEP associated with the selected endpoint minima and initial path, rather than as a proof of global uniqueness.

\section{General Overview of Defect Annihilation}
We employ the combined string and spring pair method to calculate the transformation pathway between the de-DDQC and dr-DDQC states. Due to the high dimensionality of the system, the PES features numerous local minima and saddle points, resulting in a multistage structural transformation. Nevertheless, the pathway reveals a consistent defect-annihilation mechanism, as illustrated in Fig.~\ref{fig:transformation_path}. 

Fig.~\ref{fig:transformation_path} illustrates the progressive repair of shield-like defects (red) and their induced defects (blue) throughout the path from M1 to M7. These defects are distributed in two distinct regions. The defects in the upper region A, confined within a single cluster, require only a single-stage structural transformation from M4 to M5 for elimination. Conversely, the defects in the lower region B, characterized by complex structures, require a multistage structural transformation from M1 to M7 for complete repair. As the transformation proceeds, the color-filled regions represent the defects gradually diminish, until they disappear completely. The pathway reported here differs from standard finite-temperature molecular dynamics trajectories in that it explicitly resolves the saddle points and local minima along one zero-temperature MEP. It should therefore be interpreted as a saddle-resolved energetic pathway for the selected finite cluster, rather than as a statistically sampled annealing trajectory.

 We compute the energy $V(\widetilde{\bm{r}}^N)$ of the minima and index-1 saddle points along the path from de-DDQC (M1) to dr-DDQC (M7), as shown in Fig.~\ref{fig:transformation_path}. Within the selected finite-cluster pathway, dr-DDQC has the lowest potential energy among the sampled minima and saddle points. This statement is restricted to the zero-temperature potential-energy landscape and does not establish thermodynamic stability, which would require a free-energy analysis including phason entropy and boundary effects.
 
To quantify changes in local bond-orientational order,
we define the neighbor set of particle $j$ as
\begin{equation}
\mathcal N_j
=
\left\{
k\neq j:
\|\bm r_k-\bm r_j\|_2\leq r_c
\right\},
\qquad r_c=1.1,
\end{equation}
and write $N_j=|\mathcal N_j|$.
For $N_j>0$, the local twelvefold bond-orientational
order parameter is
\begin{equation}
\phi_{12}(\bm r_j)
=
\frac{1}{N_j}
\sum_{k\in\mathcal N_j}
\exp(12\mathrm i\theta_{jk}), \label{eq:phi}
\end{equation}
where $\theta_{jk}$ is the angle of the bond
$\bm r_k-\bm r_j$ relative to a fixed reference axis.

The spatially averaged magnitude is
\begin{equation}
\overline{\Phi}_{12}
=
\frac{1}{N}
\sum_{j=1}^{N}
|\phi_{12}(\bm r_j)|.
\end{equation}
This quantity lies between zero and one and measures
local bond-angle coherence modulo $\pi/6$.
It does not retain the relative orientational phases of
different local environments and is not, by itself, a
criterion for global twelvefold quasiperiodic order.

\begin{figure}[htbp]
    \centering
    \includegraphics[width=0.48\textwidth]{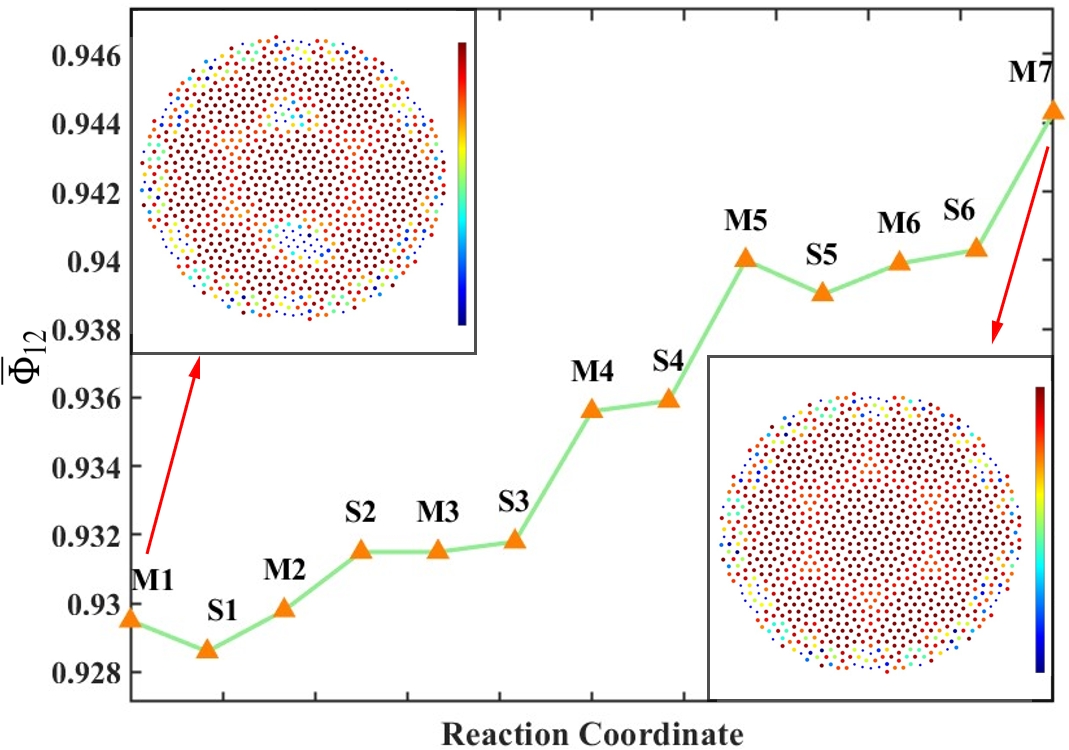}
    \caption{Pathway variation of the spatially averaged local
bond-orientational order. Mi denotes minima, Si denotes index-1 saddle points. The two insets show the spatial distribution of the local twelvefold symmetry parameter $\lvert \phi_{12} \rvert$ for the de-DDQC and dr-DDQC states, respectively.}
    \label{fig:12foldsymmetry}
\end{figure}

The endpoint values are $0.9295$ for de-DDQC and $0.9443$
for dr-DDQC, giving a net increase of $0.0148$.
Figure~\ref{fig:12foldsymmetry} shows an overall increase along the pathway,
with local decreases near some transition states.
Together with the endpoint energy difference, these results
show that the repaired reference configuration has a lower
potential energy and a higher value of this local
bond-orientational diagnostic.

\textbf{Shield-like defects}. As demonstrated previously, the transformation from de-DDQC to dr-DDQC is primarily governed by defect repair. A shield-like defect is a common type of defect in dodecagonal quasicrystals\cite{roth1998atomic,sakamoto2017defect,zhao2023molecular, gemeinhardt2018growth, zhao2025atomistic}. Its fundamental configuration is an equilateral hexagon with alternating internal angles of $\pi/2$ and $5\pi/6$ in its idealized geometry and exhibits threefold rotational symmetry\cite{fuwa2022effect}. This defect originates from a local violation or rearrangement of square--triangle matching. Its angles and side lengths relate to the square and triangular tiles, and it can be generated by slight displacements of a few atoms or by local tile flips. This geometrical compatibility allows it to be incorporated into a surrounding dodecagonal tiling and makes it a common local motif in square--triangle quasicrystal models.

\begin{figure}[htbp]
    \centering
    \includegraphics[width=0.96\columnwidth]{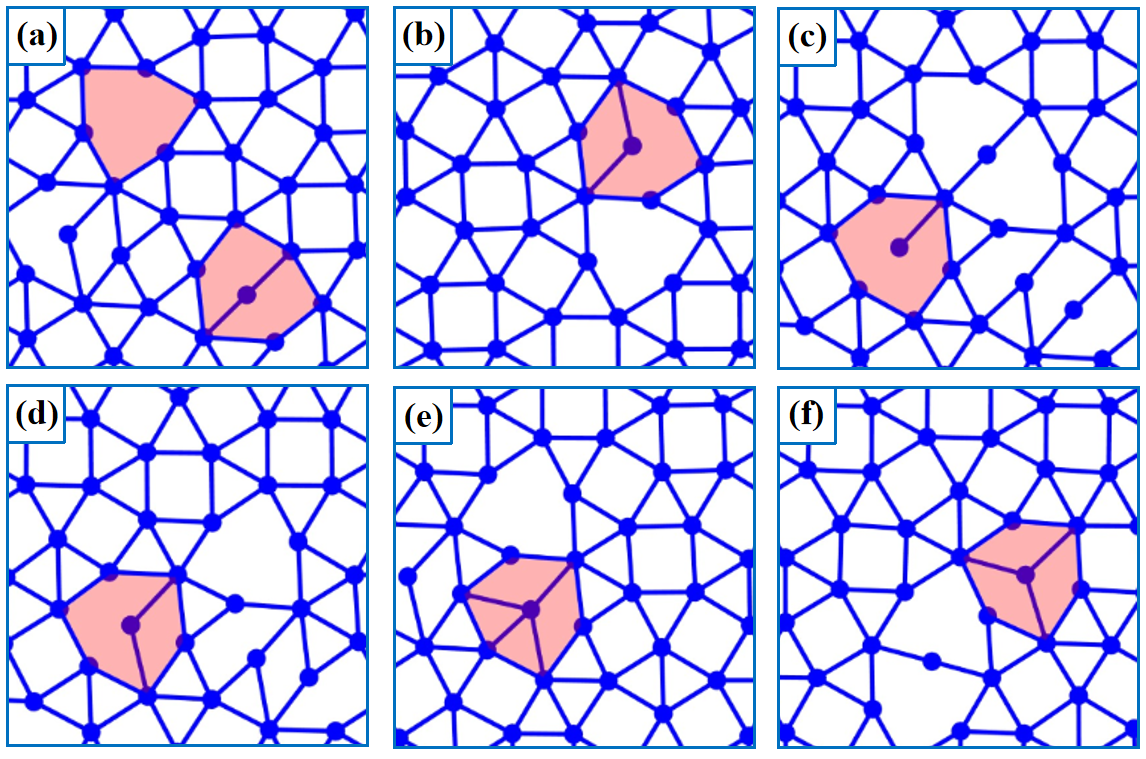}
    \caption{Types of shield-like defects. Red-filled regions represent shield-like defects.}
    \label{fig:defect_type}
\end{figure}

Analysis of de-DDQC and of the defect-repair process along the transformation pathway in Fig.~\ref{fig:transformation_path} shows that shield-like defects can be classified into different types according to the number of internal particles (0 or 1), as illustrated in Fig.~\ref{fig:defect_type}.

To quantitatively compare the stability of different types of shield-like defects, we calculate the average potential energy of the particles constituting the defect. For a shield-like defect, the set of particles it contains is denoted as $S$. We define its internal average potential energy as:
\begin{equation}
    \overline{V}_{S} = \frac{1}{N_S} \sum_{\bm{r}_j \in S} \sum_{\substack{\bm{r}_i \in S \\ \bm{r}_i \neq \bm{r}_j}} v_{\text{LJG}}(|\bm{r}_j - \bm{r}_i|). \end{equation}
where $v_{\text{LJG}}(r)$ is given by \eqref{eq:LJG} and $N_S$ denotes the number of particles in subset $S$. Using this definition, we find that the type with zero internal points, which we term the primitive hollow shield-like defect, has the lowest average energy, lying below -6. The average energies of the other types lie in the range from -6 to -5.

\section{Mechanism of Defect Annihilation}
\begin{figure}[htbp]
    \centering
    \includegraphics[width=0.96\columnwidth]{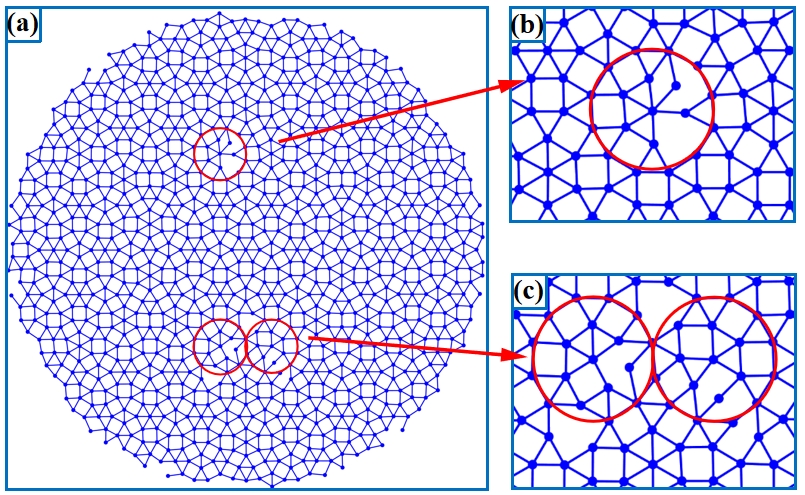}
    \caption{Distribution of shield-like defects in de-DDQC. (a) de-DDQC. The red circle and its internal area represent the local analysis region, referred to as the cluster. Panels (b) and (c) are magnified views of the corresponding regions marked in (a). (b) Cluster-1. (c) Cluster-2 and Cluster-3 from left to right. }
    \label{fig:defect_distribution}
\end{figure}

The defect distribution in the de-DDQC is complex, but along the computed MEP its local relaxation proceeds by transforming defective clusters into locally repaired square--triangle clusters. To systematically analyze the defects, we categorize the defective regions in the de-DDQC into three clusters, labeled Cluster-1 to Cluster-3, as presented in Fig.~\ref{fig:defect_distribution}.

\begin{figure*}[htbp]
    \centering
    \includegraphics[width=1.0\textwidth]{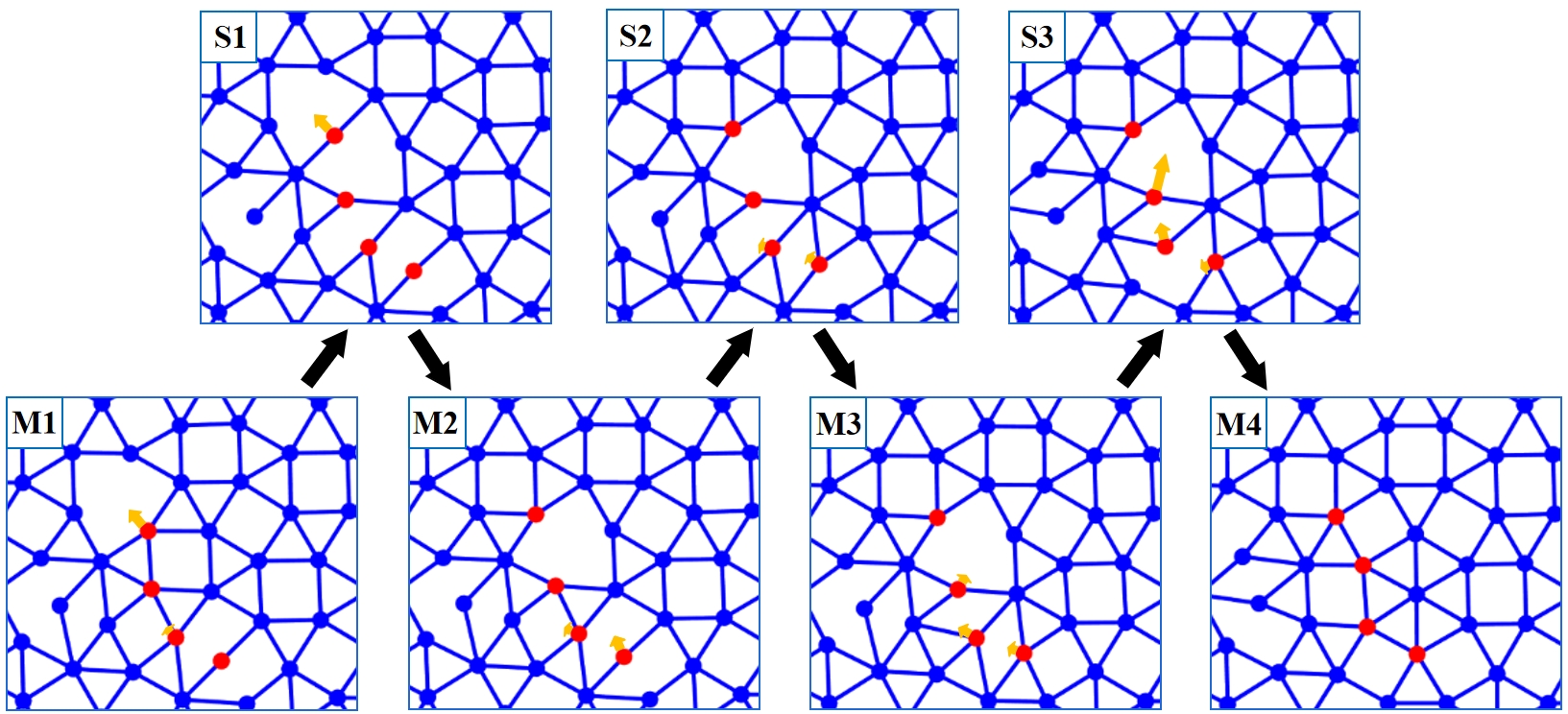}
    \caption{The defect annihilation process of Cluster-3. M1 $\rightarrow$ M2: a phason flip. M2 $\rightarrow$ M3: the aggregation of shield-like defects. M3 $\rightarrow$ M4: the decomposition of shield-like defects. Red points represent particles that undergo displacement during defect evolution, and orange arrows indicate the directions and magnitudes of their displacements. For clarity, the lengths of the orange arrows are twice the actual particle displacements.}
    \label{fig:evolution_cluster3}
\end{figure*}

Along the defect repair pathway, the system repairs the three defective clusters to form the locally repaired reference configuration. In the transformation pathway, the defect repair sequence follows Cluster-3 $\rightarrow$ Cluster-1 $\rightarrow$ Cluster-2, corresponding to the stages M1 $\rightarrow$ M4, M4 $\rightarrow$ M5, and M5 $\rightarrow$ M7, respectively in Fig.~\ref{fig:transformation_path}. The repair of Cluster-3 involves three distinct stages: phason flip (M1 $\rightarrow$ S1 $\rightarrow$ M2), aggregation (M2 $\rightarrow$ S2 $\rightarrow$ M3), and decomposition of shield-like defects (M3 $\rightarrow$ S3 $\rightarrow$ M4). Thus, we will focus primarily on the defect annihilation process in Cluster-3 in the following section. 

\subsection{The Phason Flip}
We identify the transitions $M_1\to M_2$ in Cluster-3
and $M_5\to M_6$ in Cluster-2 as phason flips from the
local tile rearrangements along the computed pathway.
For Cluster-3, the rearrangement is dominated by the
motion of one particle, accompanied by relaxation of
its surroundings. It relocates a shield-like motif from outside the
local outer ring into the interior while restoring
the outer-ring structure.

\begin{figure}[htbp]
   \centering
   \includegraphics[width=0.45\textwidth]{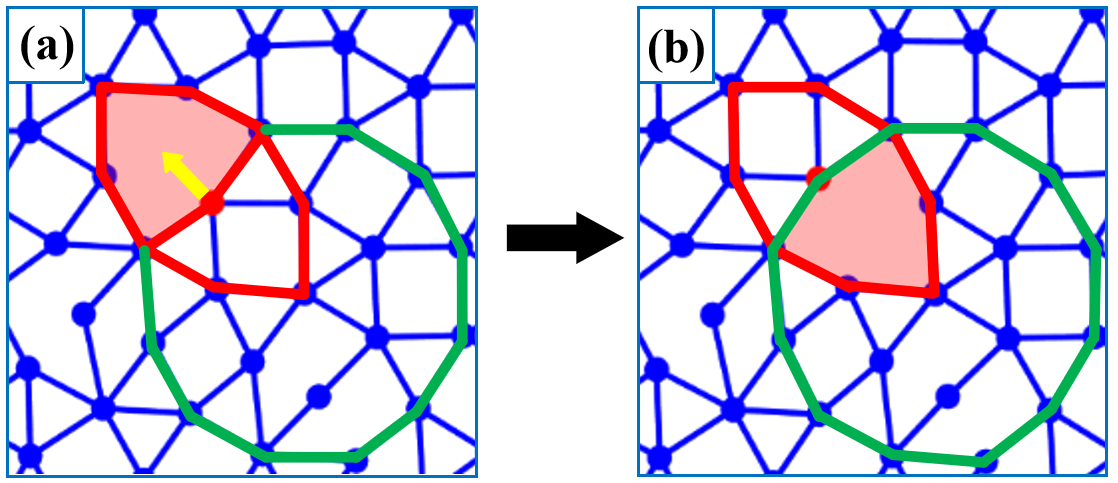}
   \caption{A phason flip. (a) Structure before the phason flip. (b) Structure after the phason flip. The red point represents the particle that undergoes a phason flip, and the yellow arrow indicates the direction and distance of the particle's movement. The area within the red outline is the region of tile changes involved in the phason flip. Green lines mark the outer ring of Cluster-3.}
   \label{fig:phasonflip}
\end{figure}

Taking Cluster-3 as an example, during the phason flip the shield-like defect, together with one square and two triangular tiles, undergoes a configurational rearrangement, as indicated within the region enclosed by the red outline in Fig.~\ref{fig:phasonflip}. This process transfers the shield-like defect from outside the cluster into its interior while simultaneously repairing the outer ring of the cluster (green outline).

\begin{figure*}[t] 
  \centering
  \includegraphics[width=0.9\linewidth]{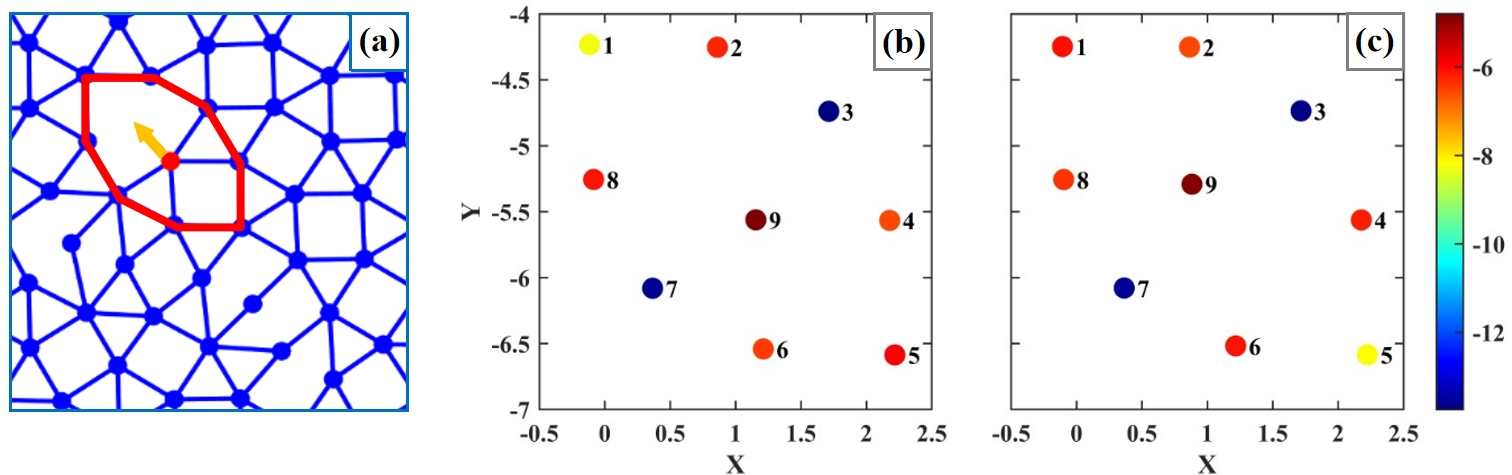}   
 \caption{Energy of particles undergoing a phason flip. (a) Structure of the phason flip in Cluster-3. (b) Potential energy distribution of the nine particles before the phason flip. (c) Potential energy distribution after the flip. The colorbar represents, for each particle, the sum of its pair interactions with the other eight particles in the selected nine-particle group.
 }
  \label{fig:energy_analysis_phason_flip} % 12pt
 \end{figure*}

 \begin{figure*}[t] 
  \centering
   \includegraphics[width=0.93\linewidth]{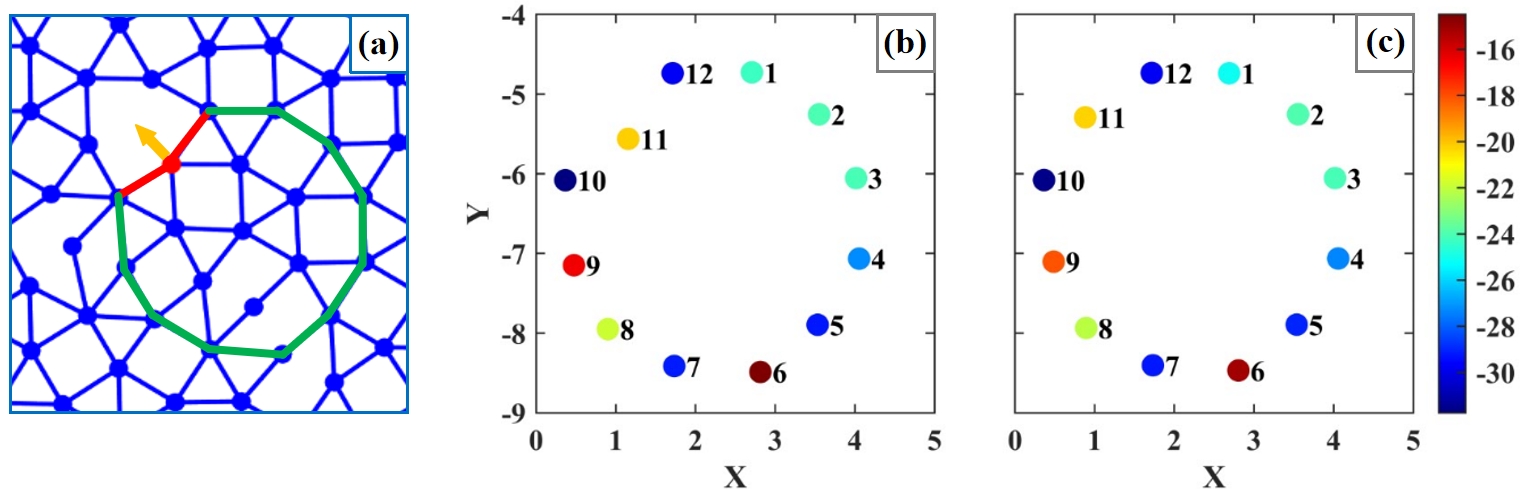}
 \caption{Energy of outer ring particles of Cluster-3. (a) Structure of Cluster-3. Green and red line segments indicate the outer ring of Cluster-3. (b) Potential energy distribution of the twelve particles on the outer ring  ($\Gamma_{\text{outer}}$) of Cluster-3 before the phason flip. (c) Potential energy distribution after the flip. }
\label{fig:energy_analysis_12fold_rotation}
\end{figure*}

From an energetic perspective, we consider only the nine particles directly involved in the phason flip in Fig.~\ref{fig:phasonflip}. We compute the sum of all pair interactions among these nine particles using the same definition as in \eqref{eq:LJG_N}, with the pair potential given by \eqref{eq:LJG}. The resulting total energies before and after the flip are -35.83 and -35.82, respectively. This energy difference of 0.01 confirms that the two configurations are nearly degenerate in energy\cite{engel2010dynamics}. This near-degeneracy is also confirmed by the total system energy. The potential energy of a particle is defined as the sum of its pair interactions with the remaining eight particles among the nine involved in the flip. Using \eqref{eq:LJG},  we find that the potential energies of these nine particles, both before and after the flip, exhibit a mirror-symmetric distribution, as shown in Fig.~\ref{fig:energy_analysis_phason_flip} (b) and (c), visually reflecting the geometric symmetry of the configurational rearrangement.
 
To quantitatively analyze the energy variation of different particle sets during defect repair, we fix the full set $\Lambda$ as the collection of all $N = 1024$ particle positions in the entire system, i.e.,
\[
\Lambda = \{\bm{r}_1, \bm{r}_2, \dots, \bm{r}_{1024}\}.
\]

For any subset $ \Gamma \subseteq \Lambda$, the average potential energy is defined as\cite{engel2007self}
\begin{equation}
    \overline{V}_{\Gamma} = \frac{1}{N_\Gamma} \sum_{\bm{r}_j \in \Gamma} \sum_{\substack{\bm{r}_i \in \Lambda \\ \bm{r}_i \neq \bm{r}_j}} v_{\text{LJG}}(|\bm{r}_j - \bm{r}_i|),
    \label{eq:V_G}
\end{equation}
where $N_\Gamma$ denotes the number of particles in subset $\Gamma$, and $v_{\text{LJG}}(r)$ is given by \eqref{eq:LJG}. This definition counts each pair within $\Gamma$ twice and each pair connecting $\Gamma$ to its complement once. Consequently, $\overline V_\Lambda=2V/N$ for the full system. We use this quantity consistently as a local interaction-energy measure along the pathway.

During the defect-repair pathway, a specific phason flip lowers the local average energy. Taking the subset $\Gamma_{\text{outer}}$ consisting of the twelve particles on the outer ring of the cluster as an example, we compute their average potential energy using \eqref{eq:V_G} and present the results in Fig.~\ref{fig:energy_analysis_12fold_rotation} (b) and (c). The flipped particle, which initially possesses a higher potential energy, moves in such a way that the average potential energy of these twelve particles decreases from -24.3813 to -24.6476 as a result of the flip, thereby promoting local repair of the cluster.

In summary, the phason flip serves two key roles. First, it repairs the outer ring of the cluster and lowers the local average energy. Second, it internalizes an external defect, thereby establishing the spatial foundation for the subsequent aggregation and decomposition of shield-like defects.

\subsection{Aggregation and Decomposition of Shield-like Defects}
\begin{figure*}[htbp]
  \centering
  \includegraphics[width=1.0\textwidth]{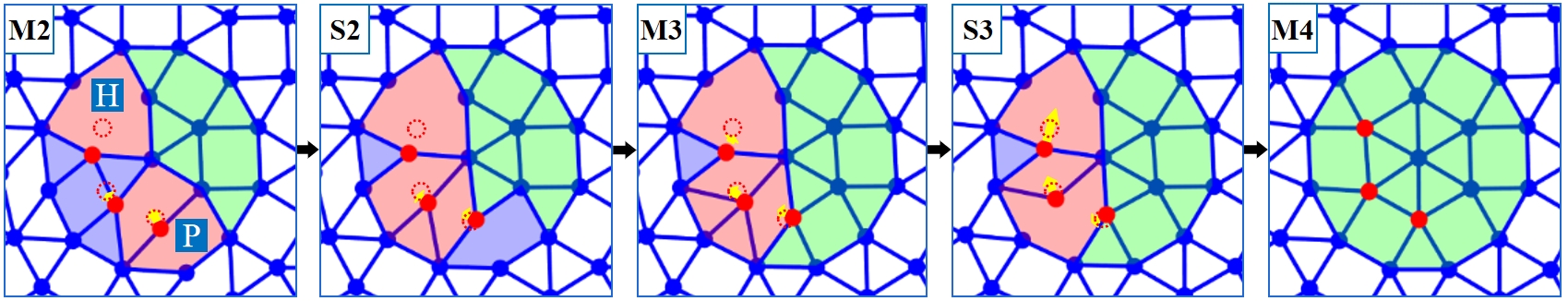}
  \caption{Aggregation and decomposition of shield-like defects. Shield-like defects are marked in red. Secondary defects that disrupt the local quasicrystalline structure are marked in blue. Locally repaired quasicrystalline regions are marked in green. The red dashed circle marks the target position in the locally repaired reference configuration for the nearest particle. H denotes the primitive hollow shield-like defect and P denotes the defect with a central particle.}
  \label{fig:aggregating_decomposition}
\end{figure*}

Both the aggregation and decomposition of shield-like defects occur inside the cluster and involve the coordinated motion of the same set of inner ring particles. Due to their highly similar driving mechanisms, these two stages are discussed together. The structural evolution for both stages is illustrated in Fig.~\ref{fig:aggregating_decomposition}.

\textbf{Aggregation.} This process refers to the merging of two separate shield-like defects via a shared edge, driven by the motion of inner ring particles. As shown in Fig.~\ref{fig:aggregating_decomposition}, we focus on the evolution of two shield-like defects marked in red, the primitive hollow defect (H) and the defect with a central particle (P). Hereafter, H and P also refer to the particle sets forming these defects. The two defects do not evolve simultaneously. Instead, the process is selectively driven by their local energies. Using \eqref{eq:V_G}, we obtain $\overline{V}_H = -25.8017$ and $\overline{V}_P = -23.1546$.
Consequently, the higher-energy defect (P) is preferentially rearranged. Its movement leads to a preliminary aggregation with the adjacent shield-like defect at the saddle point S2, eventually stabilizing at M3 via a shared edge. The aggregation provides the necessary geometric foundation for the subsequent decomposition.

\textbf{Decomposition.} The aggregated shield-like defect then decomposes into one square and two triangular tiles (M3 $\rightarrow$ M4 in Fig.~\ref{fig:aggregating_decomposition}). This stage is marked by a rapid expansion of the locally repaired (green) region. This expansion coincides with the complete removal of the local defective motif and constitutes the final step needed to fully repair the cluster.

During the aggregation and decomposition stages (M2 $\rightarrow$ M4 in Fig.~\ref{fig:aggregating_decomposition}), the repair of Cluster-3 manifests as a continuous competition: the locally repaired region expands while the defective region contracts (i.e., the green region annexes the red and blue regions). This phenomenon stems from systematic differences between the two types of regions in terms of average particle energy and symmetry order parameter. We identify the particle sets of the locally repaired region ($\Gamma_{\text{dr}}$) and the defective region ($\Gamma_{\text{de}}$) for the five structures from M2 to M4. The locally repaired region comprises green tiles, while the defective region contains red/blue shield-like defects and their induced structures.

\begin{figure*}[htbp]
  \centering
  \includegraphics[width=1.0\textwidth]{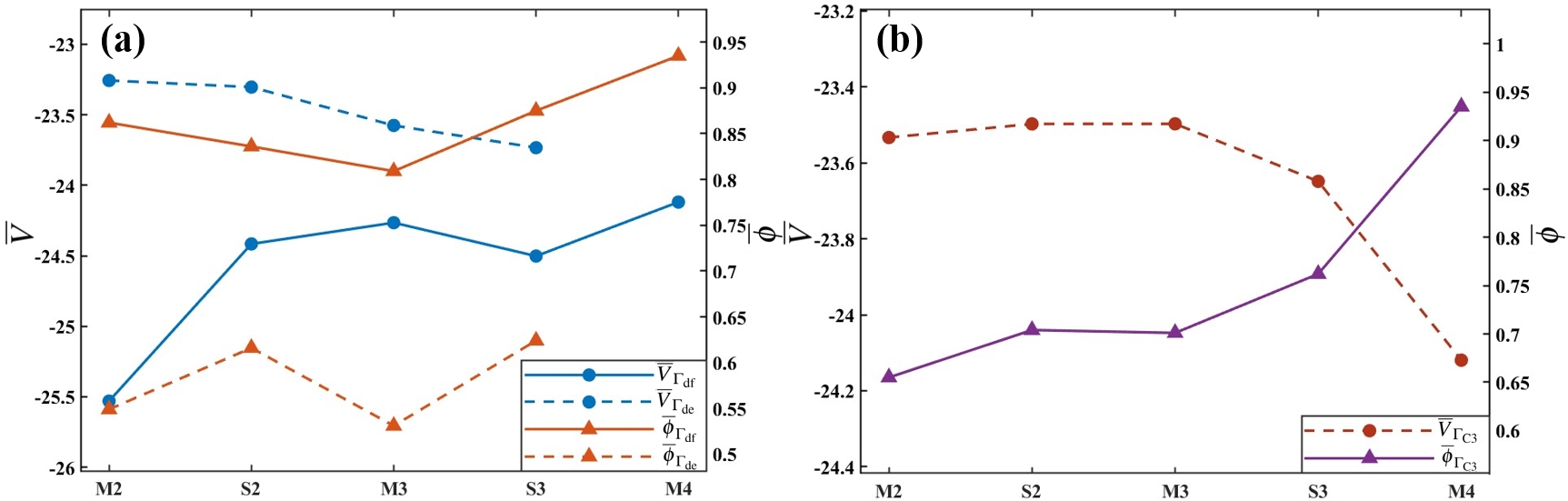}
  \caption{
Local interaction-energy and bond-orientational measures
during the aggregation--decomposition sequence
$M_2\to M_4$.
(a) Regional averages for the locally repaired region
(solid lines) and the defective region (dashed lines).
Blue curves show $\overline V_{\Gamma}$ and orange
curves show $\overline{\phi}_{\Gamma}$.
The defective-region averages are undefined at $M_4$
and are omitted.
(b) The corresponding measures for the 19 particles
of Cluster-3: $\overline V_{\Gamma_{C3}}$ in red and
$\overline{\phi}_{\Gamma_{C3}}$ in purple.
Lines connect the sampled configurations as guides
to the eye.
}
  \label{fig:energy_symmetry}
\end{figure*}

Using \eqref{eq:V_G}, we compute their average potential energies $\overline{V}_{\Gamma_{\text{dr}}}$ and $\overline{V}_{\Gamma_{\text{de}}}$. Similarly, the order parameter of average symmetry is defined as:
\begin{equation}
	\overline{\phi}_{\Gamma} = \frac{1}{N_\Gamma}\sum_{\bm{r}_j\in \Gamma}\lvert \phi_{12}(\bm{r}_j)\rvert,
	\label{eq:phi_G}
\end{equation}
where $\phi_{12}(\bm{r}_j)$ is given by \eqref{eq:phi}. Accordingly, we obtain the average symmetry parameters of the locally repaired and defective regions, $\overline{\phi}_{\Gamma_{\text{dr}}}$ and $\overline{\phi}_{\Gamma_{\text{de}}}$, respectively. These results are shown in Fig.~\ref{fig:energy_symmetry} (a). The locally repaired region consistently exhibits lower average particle energy and higher twelvefold bond-orientational order than the defective region. Therefore, along this selected MEP, the system transforms the defective region into the locally repaired structure by lowering the local potential energy, which is structurally observed as the gradual expansion of the locally repaired region.

To further analyze the impact of defect repair on the cluster, we denote the subset $\Gamma_{C3}$ as the 19 particles in Cluster-3 and analyze the changes in the average particle potential energy and symmetry. Based on the global particle positions, we compute its average potential energy $\overline{V}_{\Gamma_{C3}}$ and symmetry order parameter $\overline{\phi}_{\Gamma_{C3}}$ using \eqref{eq:V_G} and \eqref{eq:phi_G}. The evolution of these quantities is shown in Fig.~\ref{fig:energy_symmetry} (b). The data indicate that during the internal defect repair of Cluster-3, the average energy decreases while symmetry increases, and the changes in the two quantities are correlated. This abrupt change coincides with the structural rearrangement across the saddle S3 and into M4 shown in Fig.~\ref{fig:aggregating_decomposition}, confirming that this final step concentrates the most rapid energy release and structural ordering.

\section{Conclusions}
By combining the string method with the spring pair method, we have determined a zero-temperature MEP connecting a locally defective finite LJG dodecagonal quasicrystal cluster to a locally repaired reference configuration. The path reveals three elementary local rearrangements: phason flip, aggregation, and decomposition of shield-like defects. The phason flip restores the outer ring of the defect cluster through nearly degenerate single-particle motion and transports the shield-like defect into the cluster interior to enable later stages. The aggregation and decomposition stages are associated with a progressive reduction in local potential energy and an accompanying increase in local twelvefold bond-orientational order, which progressively converts defective regions into the locally repaired structure. These three stages act cooperatively along the selected path to complete the local repair of the prescribed defect combination. Our findings provide a saddle-resolved microscopic picture of one defect-repair pathway in a finite LJG dodecagonal quasicrystal cluster. The thermodynamic relevance and generality of this pathway for other potentials, boundary conditions, and defect populations remain open questions for future studies.

\begin{acknowledgments}
This work was supported by the National Key R\&D Program of China (2023YFA1008802), the Science and Technology Innovation Program of Hunan Province (2024RC1052), and the Innovative Research Group Project of National Natural Science Foundation of Hunan Province of China (2024JJ1008). We acknowledge the High-Performance Computing Platform of Xiangtan University for providing computational resources.
\end{acknowledgments}

\bibliography{reference}% Produces the bibliography via BibTeX.

\end{document}